\documentclass[aps,prb,twocolumn,notitlepage,floatfix,amsmath,amssymb]{revtex4-2}
\usepackage{graphicx,bm,booktabs}
\begin{document}
\title{Competing spiral periods and first-order transitions in layered kagome magnets}
\author{Vladimir Antropov}
\affiliation{Ames National Laboratory, Ames, Iowa 50011, USA}
\begin{abstract}
Competing interlayer exchanges can favor two spiral periods within a single classical Heisenberg Hamiltonian. For two equivalent ferromagnetic layers per crystallographic repeat, we minimize the six-exchange energy over the relative layer angle at each propagation vector. The first five exchanges cannot produce two distinct noncollinear minima of this energy, whereas exchange between equivalent layers two repeats apart allows a second minimum. We distinguish the loss of a local minimum from the equal-energy crossing at which the ground-state period changes discontinuously. The model supports at most two distinct noncollinear minima. The higher-energy spiral is unstable to out-of-plane fluctuations in the isotropic model, but sufficiently strong easy-plane anisotropy can stabilize it locally.
\end{abstract}
\maketitle
\section{Introduction}\label{sec:intro}
In a layered magnet, exchanges connecting different crystallographic repeats determine the spiral propagation vector as well as the relative angles within each repeat. When several such exchanges compete, the energy may favor more than one spiral period.

Paper I obtained the exact zero-field classical ground state of the equivalent-layer Heisenberg model using the Luttinger--Tisza method~\cite{LuttingerTisza1946,LyonsKaplan1960,KaplanMenyuk2007}, and examined its phase diagrams, harmonic stability, and magnetic-field response~\cite{PaperI}. Here we ask when the same exchanges favor two distinct noncollinear spiral periods. As the exchanges vary, each pitch changes continuously along its branch. The ground-state pitch jumps when the two branches cross at the lowest energy. Each branch describes a separate single-$q$ spiral.

In YMn$_6$Sn$_6$, a member of the layered $RT_6X_6$ (166) kagome family, the Mn moments align ferromagnetically within each layer but form incommensurate structures along the stacking direction. Neutron diffraction and field-dependent measurements reveal competing magnetic configurations~\cite{Venturini1996,Ghimire2020,Wang2021,Dally2021}. It remains unclear whether each observed period belongs to a locally stable homogeneous spiral. Electronic-structure studies of 166 compounds and the related 165 and 164 families also motivate searches for noncollinear order~\cite{Chen2026,ChenArxiv2026,XiaArxiv2026}.

Competing exchange can select a helical ground state in a classical magnet~\cite{Yoshimori}. In a frustrated $J_1$--$J_2$--$J_3$ chain, the pitch can jump at the transition from the ferromagnet to a spiral~\cite{Zinke}. Here the crossing is between two noncollinear pitches. The two-layer geometry allows us to identify the exchange required for a second interior minimum and to determine how many such minima are possible.

Rosenfeld and Mushnikov~\cite{Rosenfeld2008} described competing structures in this family using a double-flat-spiral model. We consider the six-exchange geometry of Paper I. Four exchanges join unlike layers and set their relative angle at a given wave vector. The other two connect equivalent layers and affect the preferred propagation vector. The H1 and H2 labels in Paper I describe how the layers stack. Two minima at the same exchanges can have the same stacking label.

Throughout, a ``spiral-energy minimum'' means a local minimum among homogeneous planar spirals, with the relative layer angle already optimized. We examine stability against fluctuations at other wave vectors in Sec.~\ref{sec:two}.

\section{Spiral energy and competing periods}\label{sec:energy}
\subsection{Exchange geometry and the angle between magnetic layers}
We use the six bonds of Paper I~\cite{PaperI} for two equivalent layers with rigid, equal-length spins and isotropic bilinear exchange. Layer equivalence makes the diagonal elements of the Fourier exchange matrix equal. This allows a single spiral to attain the classical ground-state energy. Unit vectors $\bm e_{An}$ and $\bm e_{Bn}$ specify the spin directions in repeat $n$. Each bond is counted once, with $J_a>0$ favoring parallel spins:
\begin{align}\label{eq:H}
\mathcal H=-\sum_n\bigl[&J_1\bm e_{An}\cdot\bm e_{Bn}
+J_2\bm e_{Bn}\cdot\bm e_{A,n+1}\nonumber\\
&+J_4\bm e_{An}\cdot\bm e_{B,n+1}
+J_5\bm e_{Bn}\cdot\bm e_{A,n+2}\nonumber\\
&+J_3(\bm e_{An}\cdot\bm e_{A,n+1}+\bm e_{Bn}\cdot\bm e_{B,n+1})\nonumber\\
&+J_6(\bm e_{An}\cdot\bm e_{A,n+2}+\bm e_{Bn}\cdot\bm e_{B,n+2})\bigr].
\end{align}

\begin{figure*}[t]
\includegraphics[width=0.90\textwidth]{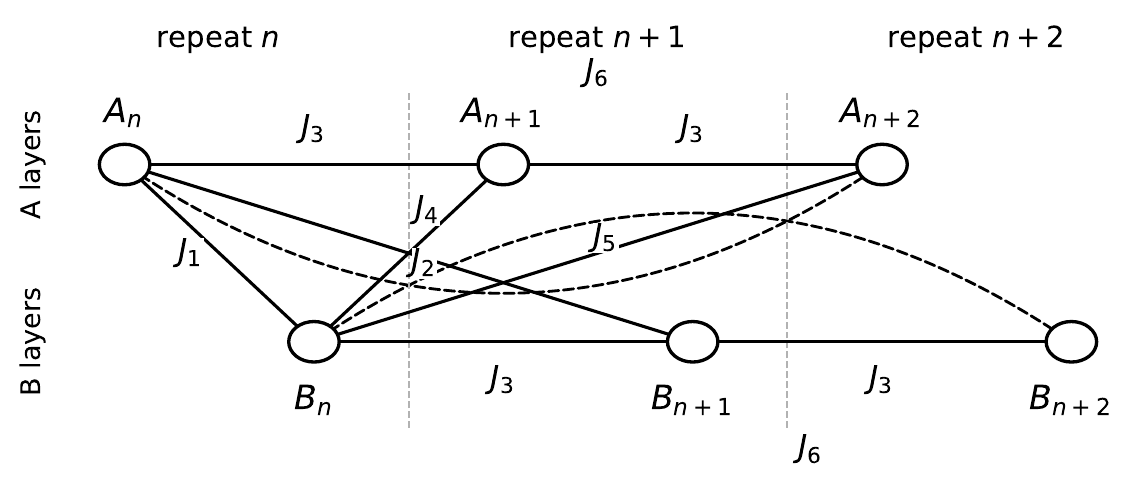}
\caption{Exchange geometry of the two-layer model. The four unlike-layer exchanges are $J_1$ ($A_n$--$B_n$), $J_2$ ($B_n$--$A_{n+1}$), $J_4$ ($A_n$--$B_{n+1}$), and $J_5$ ($B_n$--$A_{n+2}$). The equivalent-layer exchanges are $J_3$ between neighboring repeats and $J_6$ between second-neighbor repeats.}
\label{fig:geometry}
\end{figure*}
Figure~\ref{fig:geometry} shows that $J_2$ connects $B_n$ to $A_{n+1}$, while $J_4$ connects $A_n$ to $B_{n+1}$. Tracing each bond from $A$ to $B$ takes us back one repeat for $J_2$ and forward one for $J_4$. The two exchanges therefore enter the Fourier sum with opposite phases.

For a homogeneous planar spiral, let $\Phi=2\pi q$ be the turn angle between repeats and $\delta$ the relative angle between the $A$ and $B$ layers:
\begin{align}\label{eq:ansatz}
\bm e_{An}&=(\cos n\Phi,\sin n\Phi,0),\nonumber\\
\bm e_{Bn}&=(\cos(n\Phi+\delta),\sin(n\Phi+\delta),0).
\end{align}
The exchange energy per two-layer repeat is
\begin{align}\label{eq:Eraw}
E(\Phi,\delta)={}&-J_1\cos\delta-J_2\cos(\Phi-\delta)
-J_4\cos(\Phi+\delta)\nonumber\\
&-J_5\cos(2\Phi-\delta)-2J_3\cos\Phi-2J_6\cos2\Phi.
\end{align}
The four unlike-layer exchanges enter through the complex amplitude
\begin{equation}\label{eq:C}
C(\Phi)=J_1+J_2e^{-i\Phi}+J_4e^{i\Phi}+J_5e^{-2i\Phi}.
\end{equation}
For $C\ne0$, the minimum with respect to $\delta$ occurs at $e^{i\delta_*}=C^*/|C|$. The energy of the optimized spiral is
\begin{equation}\label{eq:Eopt}
\boxed{E_0(\Phi)=-2J_3\cos\Phi-2J_6\cos2\Phi-|C(\Phi)|.}
\end{equation}
At a pitch where $C=0$, the four terms in the exchange amplitude cancel. The energy is then independent of the relative layer angle $\delta$. The energy can have a cusp at such a pitch.

\subsection{Energy of the spiral states}
For a noncollinear turn angle $0<\Phi<\pi$, let $x=\cos\Phi$. The squared exchange amplitude between unlike layers is a polynomial of degree at most three in $x$
\begin{align}\label{eq:P}
P_3(x)&=|C(\Phi)|^2=a_0+a_1x+a_2x^2+a_3x^3,\\
a_0&=J_1^2+J_2^2+J_4^2+J_5^2-2(J_1J_5+J_2J_4),\nonumber\\
a_1&=2[J_1(J_2+J_4)+J_2J_5]-6J_4J_5,\nonumber\\
a_2&=4(J_1J_5+J_2J_4),\qquad a_3=8J_4J_5.\nonumber
\end{align}
The energy becomes
\begin{equation}\label{eq:Ex}
E_0(x)=-2J_3x-4J_6x^2+2J_6-\sqrt{P_3(x)}.
\end{equation}
Where $P_3(x)>0$, a stationary noncollinear spiral satisfies
\begin{equation}\label{eq:stationary}
-2J_3-8J_6x-\frac{P_3'(x)}{2\sqrt{P_3(x)}}=0.
\end{equation}
The energy cost of a small change in pitch depends on the curvature
\begin{equation}\label{eq:curvature}
E_0''(x)=-8J_6+\frac{[P_3'(x)]^2-2P_3(x)P_3''(x)}{4P_3(x)^{3/2}}.
\end{equation}
At a stationary pitch, positive curvature means that either increasing or decreasing the pitch costs energy. If the curvature vanishes, we must go to higher order to decide whether the spiral is a minimum. Two distinct minima have a maximum between them. To identify the ground state, we compare both minima with the collinear states at $x=\pm1$. For equivalent layers, the Luttinger--Tisza construction~\cite{LuttingerTisza1946,LyonsKaplan1960,KaplanMenyuk2007,PaperI} makes the absolute minimum of Eq.~\eqref{eq:Eopt} the exact zero-field classical ground-state energy of the isotropic rigid-layer model. Appendix~\ref{app:mat} gives the construction and both harmonic matrices.

\section{Why the sixth exchange is essential}\label{sec:J6}
\subsection{Why the first five exchanges cannot produce two spiral minima}
Set $J_6=0$. The spiral energy then reduces to
\begin{equation}\label{eq:E5}
E_5(x)=-2J_3x-\sqrt{P_3(x)}.
\end{equation}
To count the minima, we need the sign of the curvature. Its numerator is
\begin{equation}\label{eq:Q}
Q(x)=[P_3'(x)]^2-2P_3(x)P_3''(x).
\end{equation}
For $P_3>0$,
\begin{align}\label{eq:Qmonotone}
E_5''(x)&=\frac{Q(x)}{4P_3(x)^{3/2}},\nonumber\\
Q'(x)&=-2P_3(x)P_3'''(x)=-12a_3P_3(x).
\end{align}
Since $P_3=|C|^2\ge0$, $Q'(x)$ has a fixed sign wherever the unlike-layer exchange amplitude is nonzero. The curvature of $E_5$ can therefore change sign at most once. Two minima separated by a maximum would require two sign changes. This rules out a second noncollinear minimum on any smooth interval of pitches.

If the unlike-layer exchanges cancel at an isolated interior pitch $x=r$, the nonnegative polynomial $P_3$ has a double root there, $P_3(x)=(x-r)^2L(x)$, with $L(r)>0$. The contribution $-\sqrt{P_3}=-|x-r|\sqrt{L(x)}$ then has a downward cusp. The slope drops across $r$, excluding a minimum at the cusp. Moreover, $Q(r)=0$ and $Q$ is monotone across the interval. The curvature can thus be positive on at most one side of the cusp, where the energy is convex. If $P_3\equiv0$, the remaining energy is linear and has no isolated interior minimum.

This bound uses the bond geometry of Eq.~\eqref{eq:H}, which makes $P_3$ at most cubic in $x$.

\subsection{The physical role of $J_6$}
The $J_6$ interaction couples equivalent layers two crystallographic repeats apart. Its contribution, $-2J_6\cos2\Phi=-4J_6x^2+2J_6$, changes the second derivative by $-8J_6$ without changing the optimum relative angle $\delta$ at fixed $\Phi$:
\begin{equation}\label{eq:J6curv}
E_0''(x)=E_5''(x)-8J_6.
\end{equation}
The four unlike-layer exchanges determine how the curvature of $E_5$ varies with pitch. The shift due to $J_6$ can give the energy two regions of positive curvature. Changing $J_3$ shifts the slope by the same amount at every $x$. A minimum occurs wherever the slope passes through zero from below. Varying $J_3$ or $J_6$ can reverse the energy ordering of the two minima or cause one minimum to merge with the maximum between them.

\section{Two competing spiral periods}\label{sec:two}
\subsection{Two minima near an energy crossing}
The following illustrative exchange constants, in units of $J_1$, produce two homogeneous spiral minima:
\begin{align}\label{eq:example}
(J_1,J_2,J_3)/J_1&=(1,0.140991,-0.140615),\nonumber\\
(J_4,J_5,J_6)/J_1&=(0.05137,0.333651,-0.249118).
\end{align}
Rounding the exchanges to three decimal places changes the energy difference between the two spirals.

\begin{figure}[t]
\includegraphics[width=\columnwidth]{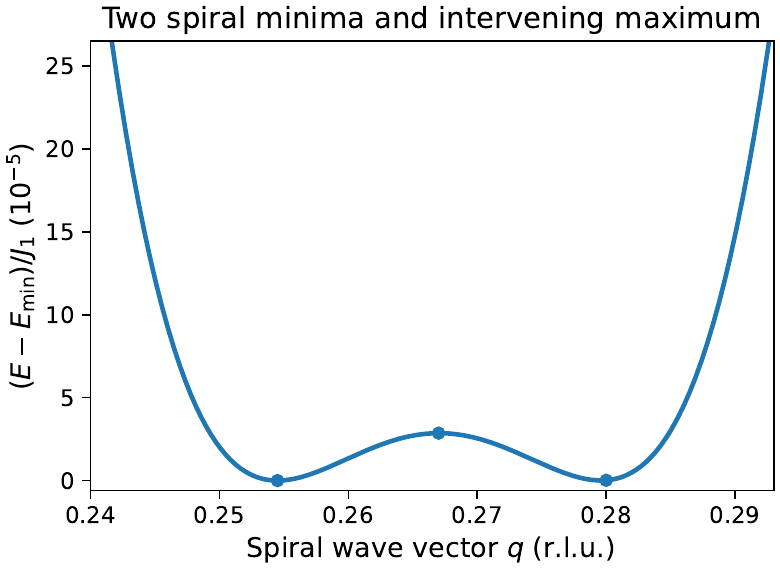}
\caption{Spiral energy for the exchanges in Eq.~\eqref{eq:example}, measured from the lower minimum. The vertical scale is $10^{-5}J_1$. The splitting between the minima is given in Table~\ref{tab:branches}. The maximum gives the barrier along the path of homogeneous spirals.}\label{fig:energy}
\end{figure}

The two minima and the maximum between them are visible in Fig.~\ref{fig:energy}. Table~\ref{tab:branches} gives their positions and curvatures. The minima differ by $6.25\times10^{-8}J_1$ per two-layer repeat, while the intervening maximum lies about $2.86\times10^{-5}J_1$ above the lower minimum. The small splitting makes the preferred period sensitive to changes in the exchanges.
\begin{table}[t]
\caption{Homogeneous planar spiral solutions for Eq.~\eqref{eq:example}. The second derivatives refer to $x=\cos(2\pi q)$, and energies are measured from the lower minimum.}\label{tab:branches}
\begin{ruledtabular}\begin{tabular}{lccc}
Branch & $q$ (r.l.u.) & $(E_0-E_{\min})/J_1$ & $E_0''(x)/J_1$\\
Low-$q$ minimum & 0.255 & 0 & 0.037\\
Intermediate maximum & 0.267 & $2.86\times10^{-5}$ & $-0.019$\\
High-$q$ minimum & 0.280 & $6.25\times10^{-8}$ & 0.037
\end{tabular}\end{ruledtabular}
\end{table}

\subsection{Harmonic stability of the two spirals}
A minimum of $E_0(q)$ is stable against uniform changes of pitch, but may still be unstable against fluctuations at other wave vectors. For the unit spins used here, an out-of-plane fluctuation changes the energy per repeat by $\Delta E_\perp=\frac12\sum_p\bm v(p)^\dagger\mathsf H_\perp(p)\bm v(p)$. The lower eigenvalue is~\cite{PaperI}
\begin{equation}\label{eq:perpPaperI}
\lambda_{\perp,-}(p)=E_0(p)-E_0(\Phi).
\end{equation}
Here $\Phi$ is the turn angle of the reference spiral and $p$ is the fluctuation wave number. Both are in radians per crystallographic repeat. If $E_0(\Phi_1)<E_0(\Phi_2)$, the out-of-plane mode of spiral 2 at $p=\Phi_1$ has eigenvalue
\begin{equation}\label{eq:unstablePaperI}
\lambda_{\perp,-}(\Phi_1)=E_0(\Phi_1)-E_0(\Phi_2)<0.
\end{equation}
The higher-energy spiral can lower its energy by tilting out of plane at the wave vector of the competing spiral. When the spiral energies coincide, the quadratic energy cost of this tilt vanishes [Eq.~\eqref{eq:unstablePaperI}]. If their common energy is the absolute minimum of $E_0(p)$, Paper I establishes both single-$q$ spirals as exact classical ground states. To decide whether a continuous family of spirals or a stable double-$q$ state shares this energy, we would need to go beyond quadratic order. Other interactions may select among degenerate configurations.

Consider the easy-plane anisotropy used in Paper I:
\begin{equation}\label{eq:easyplane}
\mathcal H_K=K\sum_{n,\alpha=A,B}(e^z_{\alpha n})^2,\qquad K>0,
\end{equation}
so the energy of a planar spiral is unchanged, while every out-of-plane harmonic eigenvalue increases by $2K$ in the normalization of Eq.~\eqref{eq:perpPaperI}. The higher-energy spiral has no negative out-of-plane eigenvalue when
\begin{equation}\label{eq:Kthreshold}
K\geq K_c=\frac{E_0(\Phi_2)-E_0(\Phi_1)}{2}.
\end{equation}
For Eq.~\eqref{eq:example}, the splitting $\Delta E=6.25\times10^{-8}J_1$ gives $K_c\simeq3.13\times10^{-8}J_1$. The competing out-of-plane mode has positive quadratic energy for $K>K_c$.

A uniform rotation costs no energy. For both spirals the calculated in-plane spectrum is positive at every other sampled wave vector. With $p$ in radians per repeat, the long-wavelength behavior of the lower branch is $\lambda_{\parallel,-}(p)=\rho p^2+O(p^4)$. At a stationary pitch, $\rho=\tfrac12\,d^2E_0/d\Phi^2$ in the normalization of Appendix~\ref{app:mat}. This gives $\rho/J_1=0.0183$ and $0.0180$ for the low- and high-$q$ spirals, respectively. For these exchanges, $K>K_c$ makes the higher-energy spiral locally stable at harmonic order. We have not calculated its lifetime or the barrier along other paths between the spirals.

\section{Competition and first-order transitions between spiral periods}\label{sec:boundaries}
\subsection{Appearance and disappearance of spiral minima}
To find where two spiral minima exist, we vary $J_3$ and $J_6$ while holding $J_1,J_2,J_4,J_5$ fixed. Writing $F(x)=-\sqrt{P_3(x)}$,
\begin{equation}\label{eq:foldenergy}
E_0(x)=F(x)-2J_3x-4J_6x^2+2J_6.
\end{equation}
At the boundary of this region, a minimum merges with the adjacent maximum, so both the first and second derivatives vanish:
\begin{align}
J_6^{\rm fold}(x)&=\frac{F''(x)}{8},\label{eq:fold6}\\
J_3^{\rm fold}(x)&=\frac{F'(x)-xF''(x)}{2},\label{eq:fold3}
\end{align}
with $-1<x<1$ and $P_3(x)>0$. If the higher-energy minimum disappears, the other spiral remains the ground state provided its energy also lies below both collinear states.

\begin{figure}[t]
\includegraphics[width=\columnwidth]{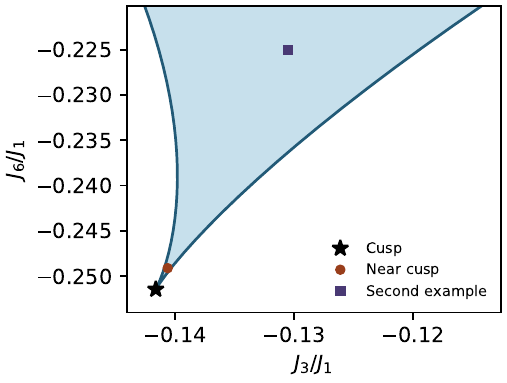}
\caption{For exchanges in the shaded region of the $(J_3,J_6)$ plane, the spiral energy has two local minima as a function of pitch. The other four exchanges have the values in Eq.~\eqref{eq:example}. The star marks the cusp where the two boundaries meet, the square the second example, and the circle the exchanges of Fig.~\ref{fig:energy}. The boundaries correspond to the appearance or disappearance of a local minimum, not necessarily to a ground-state transition.}\label{fig:folds}
\end{figure}

Where the two boundaries meet, $F'''(x_c)=0$ as well. For the four unlike-layer exchanges in Eq.~\eqref{eq:example},
\begin{equation}\label{eq:cusp}
x_c=-0.107,\quad
J_{3c}/J_1=-0.142,\quad
J_{6c}/J_1=-0.251.
\end{equation}
At the cusp, the two minima and the maximum between them merge at $q_c\simeq0.267$. The exchanges in Eq.~\eqref{eq:example} lie close to this cusp. Farther inside the two-minimum region, at $(J_3,J_6)/J_1=(-0.1305,-0.225)$, the minima occur at $q=0.221$ and $0.313$. The maximum at $q=0.267$ lies $4.05\times10^{-3}J_1$ per repeat above the lower minimum, about 140 times the uniform-pitch barrier in Fig.~\ref{fig:energy}. The two examples lie inside the boundaries in Fig.~\ref{fig:folds}.

\subsection{Equal energies and the ground-state transition}
A discontinuous change of the ground-state period occurs when two distinct spiral minima reach the same lowest energy. For interior minima $x_1\ne x_2$, write $s(x)=\sqrt{P_3(x)}$ and $g(x)=s'(x)$. Their stationary pitches satisfy
\begin{equation}\label{eq:coexstat}
2J_3+8J_6x_i+g(x_i)=0,\qquad i=1,2.
\end{equation}
Equating the energies, $E_0(x_1)=E_0(x_2)$, gives
\begin{equation}\label{eq:secant}
\frac{s(x_2)-s(x_1)}{x_2-x_1}=\frac{g(x_1)+g(x_2)}{2}.
\end{equation}
The exchange constants at the energy crossing then satisfy
\begin{align}
J_6&=-\frac{g(x_2)-g(x_1)}{8(x_2-x_1)},\label{eq:coexJ6}\\
J_3&=-\frac{g(x_1)}{2}-4J_6x_1.\label{eq:coexJ3}
\end{align}

With $J_1,J_2,J_4,J_5,J_6$ fixed at the values in Eq.~\eqref{eq:example}, equal spiral energies occur at
\begin{equation}\label{eq:coexnumber}
J_3^{\rm coex}/J_1\simeq-0.1406152,
\end{equation}
with propagation vectors $q_1\simeq0.255$ and $q_2\simeq0.280$. Their common energy lies below the intervening maximum and both collinear endpoints. Since Sec.~\ref{sec:no3ss} excludes a third noncollinear minimum, these are the absolute minima of $E_0$. The construction in Appendix~\ref{app:mat} then gives an exact classical ground state at each pitch. At the crossing, the out-of-plane mode at the competing wave vector costs no energy to second order. On either side it lowers the energy of the higher-energy spiral [Eq.~\eqref{eq:unstablePaperI}].

\begin{figure}[t]
\includegraphics[width=\columnwidth]{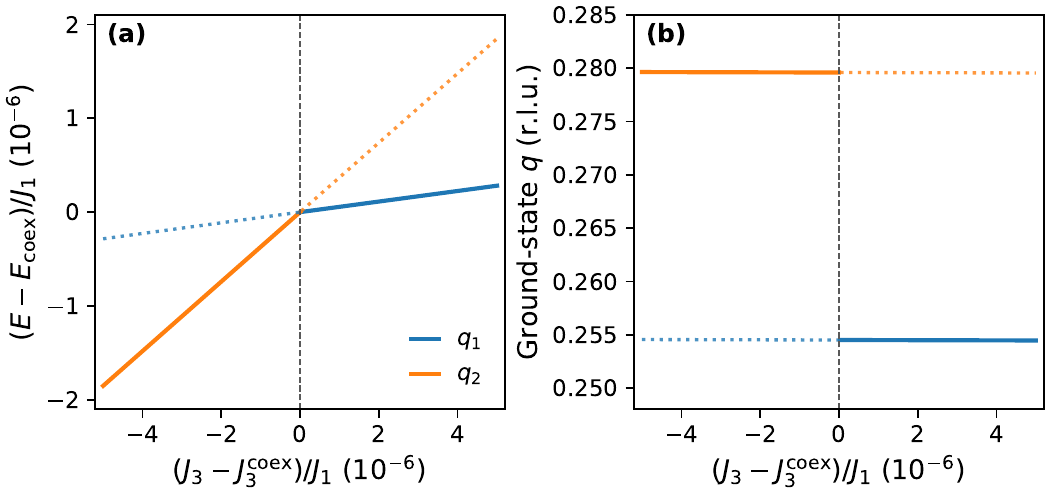}
\caption{Change of the ground-state spiral period as $J_3$ passes through coexistence, with the other five exchanges fixed at Eq.~\eqref{eq:example}. (a) Energies of the two stationary spirals measured from their common energy at coexistence. The horizontal scale is $(J_3-J_3^{\rm coex})/J_1$ in units of $10^{-6}$. (b) Ground-state wave vector (solid lines), with dotted continuations of the higher-energy branches. The dashed line marks the crossing at which the preferred wave vector jumps from $q_1\simeq0.255$ to $q_2\simeq0.280$.}\label{fig:transition}
\end{figure}

The pitch of each spiral varies continuously with $J_3$. When their energies cross, the ground-state wave vector jumps by $\Delta q\simeq0.025$ r.l.u. (Fig.~\ref{fig:transition}).

\section{Why there cannot be three competing spiral minima}\label{sec:no3ss}
For the bonds of Eq.~\eqref{eq:H}, no choice of exchange signs or strengths gives more than two distinct noncollinear minima of the optimized homogeneous spiral energy.

The two equivalent-layer exchanges add only linear and quadratic terms in $x=\cos\Phi$. The third derivative of $E_0(x)$ therefore comes entirely from the unlike-layer contribution $-\sqrt{P_3(x)}$, where $P_3\ge0$ is at most cubic:
\begin{align}\label{eq:third}
E_0'''(x)&=-\frac{R(x)}{8P_3(x)^{5/2}},\nonumber\\
R&=4P_3^2P_3'''-6P_3P_3'P_3''+3(P_3')^3.
\end{align}
On each interval where $P_3>0$, $R(x)$ has at most one distinct zero, except when $\sqrt{P_3}$ is polynomial and the energy is quadratic on the smooth interval. Three minima would require two intervening maxima and hence at least three zeros of $E_0'''$, which Eq.~\eqref{eq:third} excludes. A zero of $C$ produces a downward cusp, not another minimum. Appendix~\ref{app:count} gives the proof, including the zero-amplitude cases.

The count covers only interior minima of the homogeneous spiral energy. Either collinear endpoint can still be the ground state.

\section{Testing calculated interlayer exchanges}\label{sec:materials}
Comparing a ferromagnetic state with a few antiferromagnetic configurations can reveal competing exchanges, but does not establish whether a noncollinear spiral has lower energy or whether two spiral minima exist. For candidate 166 kagome compounds and the related 165 and 164 families~\cite{Chen2026,ChenArxiv2026,XiaArxiv2026}, the spiral energy must be calculated for the layer connectivity of each structure.

Each of the six layer exchanges in Eq.~\eqref{eq:H} is a sum of microscopic exchange constants. Each bond type contributes its exchange constant times the number of such bonds per repeat connecting the two layers. Equation~\eqref{eq:Eopt} then gives the energy as a function of pitch. We compare the energies of stationary spirals in $0<q<1/2$ with those at zeros of $C$ and at the collinear endpoints. Squaring Eq.~\eqref{eq:stationary} can introduce extra roots, so we keep only those that satisfy the original equation. The lowest energy identifies the ground state within the rigid-layer model.

For two minima, $\Delta q=|q_2-q_1|$ measures the separation of the preferred wave vectors and $\Delta E=E_0(q_2)-E_0(q_1)$ their energy splitting. The curvature at each minimum gives the cost of a small change in pitch. The height of the intervening maximum gives the barrier when the pitch changes uniformly throughout the crystal. Changing the pressure or composition could tune the interlayer exchanges through the crossing in Sec.~\ref{sec:boundaries}. The preferred period would then switch while both minima survive.

For a higher-energy spiral, Eq.~\eqref{eq:unstablePaperI} identifies its unstable out-of-plane mode in the isotropic model, and Eq.~\eqref{eq:Kthreshold} gives the anisotropy required to remove it. Additional magnetic sublattices, variable moment magnitudes, or other interactions require a different energy and stability analysis. If a calculation gives three distinct interior minima, we must check whether the material satisfies the assumptions of Eq.~\eqref{eq:H} and whether the calculated exchanges have been assigned to the correct interlayer bonds.

\section{Relation to experiments on YMn$_6$Sn$_6$}\label{sec:ymn}
Neutron diffraction on YMn$_6$Sn$_6$ shows two nearby incommensurate propagation vectors. They approach one another on cooling and both persist to low temperature~\cite{Ghimire2020,Dally2021}. We have not fitted the exchanges of Eq.~\eqref{eq:example} or the anisotropy to this compound, so we make no assignment of the model wave vectors $q\simeq0.255$ and $0.280$ to the measured peaks. In the model, easy-plane anisotropy can stabilize two spirals with different exchange energies.

Competing diffraction peaks and field-induced structures~\cite{Ghimire2020,Wang2021,Dally2021} can arise from homogeneous spiral phases or from nonuniform modulations. Spin-polarized neutron diffraction shows that an electric current and a magnetic field can control the handedness of the spiral in YMn$_6$Sn$_6$~\cite{Masuda2026}. Spirals of opposite handedness can have the same period, so this measurement leaves open whether two equilibrium periods coexist.

At the equal-energy crossing in Fig.~\ref{fig:transition}, both pitches minimize the exchange energy and domains of either spiral may occur. The relative peak intensities depend on how much of the sample adopts each pitch. Domain walls, disorder, and other nonuniform spin configurations can affect those fractions. Equation~\eqref{eq:Kthreshold} gives the easy-plane anisotropy needed to remove the out-of-plane instability of the higher-energy spiral. The domain populations remain undetermined.

A weak field in the spiral plane lowers the energy by distorting the angles between spins. For each locally stable planar spiral the relaxed energy is
\begin{equation}\label{eq:fieldcross}
E_i(h)=E_i(0)-\frac12\chi_{\parallel,i}h^2+O(h^4).
\end{equation}
If the higher-energy spiral has the larger in-plane susceptibility, the field lowers its energy more rapidly. To quadratic order, the two energies cross at
\begin{equation}\label{eq:hstar}
h_*\simeq\sqrt{\frac{2\Delta E}{\chi_{\parallel,2}-\chi_{\parallel,1}}}.
\end{equation}
We keep the spins in the plane and use the exchanges in Eq.~\eqref{eq:example}. The matrix in Appendix~\ref{app:mat} then gives the susceptibilities of the lower-energy spiral (1) and the higher-energy spiral (2),
\begin{equation}\label{eq:chivalues}
\chi_{\parallel,1}=2.937/J_1,\qquad
\chi_{\parallel,2}=3.113/J_1,
\end{equation}
where both susceptibilities are per two-layer repeat. Using $\Delta\chi_\parallel\simeq0.176/J_1$ and $\Delta E\simeq6.25\times10^{-8}J_1$ gives
\begin{equation}\label{eq:hstarvalue}
h_*/J_1\simeq8.42\times10^{-4}.
\end{equation}
Both are stable against in-plane fluctuations (Sec.~\ref{sec:two}). Stability of the higher-energy spiral against out-of-plane tilts also requires sufficient easy-plane anisotropy.

To test the weak-field expansion, we minimized the classical exchange and Zeeman energies in periodic cells containing $10^4$ two-layer repeats, allowing independent planar angles on both layers. The allowed pitches $q_1=0.2545$ and $q_2=0.2796$ r.l.u. agree with the continuous minima to the precision shown. For a pure spiral of winding $k$ in $N$ repeats, we initialized the $A$-layer angles with a constant increment $2\pi k/N$ and set the relative layer angle to $\delta_*(2\pi k/N)$. Taking the field along $x$, we relaxed all $2N$ angles with L-BFGS-B using the analytic derivatives of the exchange and Zeeman energies. Both layers retained their initial winding after relaxation.

The relaxed energies cross between $h/J_1=8.42\times10^{-4}$ and $8.45\times10^{-4}$, consistent with Eq.~\eqref{eq:hstarvalue}. Across this interval, $[E_2(h)-E_1(h)]/J_1$ changes from $+3.2\times10^{-11}$ to $-4.1\times10^{-10}$. To establish a phase boundary, we would also need to compare other winding numbers and states with domains of different pitch.

To compare pure and mixed periods, we used cells of 400 two-layer repeats at $h/J_1=8.4\times10^{-4}$ and $10^{-3}$, scanning $0.245\leq q\leq0.285$. The scan resolves energy minima near both spiral pitches. For the mixed initial state, the first 200 increments of the $A$-layer angle were $2\pi(0.255)$ and the next 200 were $2\pi(0.280)$, giving $q=107/400=0.2675$. In each half we used the relative layer angle of the corresponding pure spiral and then relaxed all angles at each field.

For the 400-repeat comparison we measure energies per repeat from the continuous zero-field minimum, $E_{\min}/J_1=-1.17060497826$. At $h/J_1=8.4\times10^{-4}$ and $10^{-3}$, the lowest sampled pure spiral has $q=112/400=0.280$, with $(E-E_{\min})/J_1=-0.92\times10^{-6}$ and $-1.38\times10^{-6}$, respectively. For the relaxed mixed state, the energy lies above the zero-field minimum by $3.78\times10^{-6}J_1$ at the lower field and $3.33\times10^{-6}J_1$ at the higher field. The mixed configurations lie about $4.7\times10^{-6}J_1$ per repeat above the lowest sampled pure-period states, although their energy is below that of a homogeneous spiral constrained to the same intermediate winding. Here we tested one arrangement of the two regions in the 400-repeat cell.

On the 400-repeat grid at zero field, the $q=0.255$ spiral lies only $7.1\times10^{-9}J_1$ per repeat below the $q=0.280$ spiral. The spacing $1/400$ reduces the zero-field splitting by nearly a factor of nine relative to the continuous pitches. The high-$q$ state is therefore lower at both sampled fields, and this grid cannot locate the crossing of the continuous branches.

A field normal to the spiral plane instead produces conical spirals in the isotropic model. Their saturation fields are nearly equal, $h_{{\rm sat},1}/J_1\simeq h_{{\rm sat},2}/J_1\simeq0.4241$, with $h_{{\rm sat},1}-h_{{\rm sat},2}\simeq6.25\times10^{-8}J_1$. Below saturation, their energies satisfy $E_i(h)=E_i(0)-h^2/h_{{\rm sat},i}$, giving
\begin{equation}\label{eq:hardaxisdifference}
E_2(h)-E_1(h)=\Delta E\left[1-\frac{h^2}{h_{{\rm sat},1}h_{{\rm sat},2}}\right]>0,
\end{equation}
for $0\le h<\min(h_{{\rm sat},1},h_{{\rm sat},2})$. With the field normal to the spiral plane, the spiral favored at zero field stays lower in energy until the first of the two spirals saturates. Material-specific anisotropy and domains can alter this behavior.

The full exchange fit for YMn$_6$Sn$_6$ at $U-J=0.4$ eV in Supplemental Table S2 of Ref.~\cite{Ghimire2020} gives $J_6/J_1=0.54/(-12.86)\simeq-0.042$. Their angles $\alpha$ and $\beta$ are the successive turns across the two spacers. The term $2J_6\cos(2\alpha+2\beta)$ in their Supplemental Eq.~(S1) couples layers separated by two repeats, since $\alpha+\beta=\Phi$. Reversing the exchange sign convention leaves the ratio unchanged. Relative to $J_1$, this antiferromagnetic exchange is about one-fifth to one-sixth as large as in our two examples.

A separate first-principles study~\cite{Sadhukhan} retains five effective interlayer exchanges. Its microscopic $J_6$ enters the sum for a nearest-layer coupling, so its tabulated pair exchanges do not give the ratio in Eq.~\eqref{eq:H}. Whether another 166 compound has an antiferromagnetic $J_6$ of comparable strength relative to $J_1$ remains open.

One could test this interpretation in YMn$_6$Sn$_6$ by determining its interlayer exchanges and following the diffraction peaks as pressure or composition changes. A discontinuous period change near an equal-energy crossing would support competition between two exchange-selected pitches.

\section{Discussion and conclusions}\label{sec:conclusions}
For the two-layer geometry of Eq.~\eqref{eq:H}, $J_1$--$J_5$ cannot produce two distinct noncollinear spiral-energy minima. Coupling $J_6$ between equivalent layers two repeats apart can produce a second minimum, but the six-exchange model admits no third noncollinear minimum.

A spiral minimum disappears when it merges with the intervening maximum. At the first-order transition considered here, both minima survive and the preferred pitch switches when their energies cross. When the two spirals share the lowest energy, each is an exact classical ground state of the isotropic model. An out-of-plane tilt at the competing wave vector then costs no energy to second order. Away from that crossing, the higher-energy spiral requires sufficient easy-plane anisotropy for local stability.

A change of preferred pitch need not change the H1/H2 stacking label of Paper I. The illustrative exchange set has $|J_6|/J_1\simeq0.249$, but Fig.~\ref{fig:folds} shows that the range supporting two minima depends on all the other exchanges.

At finite temperature, the difference in spin-wave free energy between the two spirals could compete with their small exchange-energy splitting. We have not calculated this thermal contribution.

\appendix
\section{Ground-state energy and harmonic matrices}\label{app:mat}
For Fourier amplitudes proportional to $e^{ipn}$, write $d(p)=2J_3\cos p+2J_6\cos2p$. The exchange matrix has diagonal entries $d(p)$ and upper off-diagonal entry $C(p)$ from Eq.~\eqref{eq:C}. Its largest eigenvalue at $p$ is $d(p)+|C(p)|=-E_0(p)$. For this mode we can choose equal spin amplitudes on the two layers. For nonzero $C$, the phase of layer $B$ relative to layer $A$ is $-\arg C(p)$. When $C=0$, any relative phase gives the same energy. At the wave vector where the exchange eigenvalue is largest, the planar spiral has unit spins on both layers and reaches the Fourier lower bound on the energy.

Let $\Phi$ be the reference pitch, $\Lambda=-E_0(\Phi)$, and $\delta=\delta_*(\Phi)$. For either an out-of-plane tilt or an in-plane rotation, the two components of $\bm v_n$ are the fluctuations on layers $A$ and $B$ in repeat $n$. Their quadratic energy is $\frac12\sum_p\bm v^\dagger\mathsf H\bm v$ per repeat, with $\bm v(p)=N^{-1}\sum_n e^{-ipn}\bm v_n$ for $N$ repeats. In zero field the matrices are
\begin{align}
\mathsf H_\perp(p)&=\begin{pmatrix}\Lambda-d(p)&-C(p)\\-C(p)^*&\Lambda-d(p)\end{pmatrix},\\
\mathsf H_\parallel(p)&=\begin{pmatrix}\Lambda-d_\Phi(p)&-B_\Phi(p)\\-B_\Phi(p)^*&\Lambda-d_\Phi(p)\end{pmatrix}.
\end{align}
For an in-plane rotation, each bond contributes its exchange constant multiplied by the cosine of the angle between the two spins in the reference spiral. With $r=(0,-1,1,-2)$ and corresponding $j_r=(J_1,J_2,J_4,J_5)$, this gives
\begin{align}
B_\Phi(p)&=\sum_r j_r\cos(\delta+r\Phi)e^{irp},\\
d_\Phi(p)&=2J_3\cos\Phi\cos p+2J_6\cos2\Phi\cos2p .
\end{align}
The lower in-plane eigenvalue is $\Lambda-d_\Phi(p)-|B_\Phi(p)|$. Its expansion at $p=0$ includes relaxation of the relative layer angle, as does the pitch derivative of the optimized energy in Eq.~\eqref{eq:Eopt}.

\section{Maximum of two spiral-energy minima}\label{app:count}
To count the interior minima, we write the spiral energy as
\begin{equation}\label{eq:appE}
E(x)=\alpha x+\beta x^2-\sqrt{P(x)}+\mathrm{const},\qquad -1<x<1,
\end{equation}
where $P(x)=a+bx+cx^2+dx^3\ge0$ is the squared interlayer exchange amplitude. The terms linear and quadratic in $x$ do not contribute to the third derivative, so Eq.~\eqref{eq:third} applies with $P$ replacing $P_3$ wherever $P>0$.

\paragraph{Cubic polynomial.} First consider a cubic ($d\ne0$) with no repeated roots. It has either one or three real roots. A shift of $x$ removes its quadratic term, and rescaling (with reversal if needed) gives $P(t)=t^3+pt+q$. The numerator of $E'''$ is
\begin{align}\label{eq:appR}
R(t)=3\bigl[&p^3+5p^2t^2+4pqt-5pt^4\nonumber\\
&+8q^2-20qt^3-t^6\bigr].
\end{align}
The discriminant of this polynomial is
\begin{equation}\label{eq:appdisc}
\operatorname{Disc}_t R=2^{18}3^{10}(4p^3+27q^2)^5.
\end{equation}
The factor $4p^3+27q^2$ vanishes when $P$ has a repeated root. In the other cases, the number of real zeros of $R$ remains unchanged because its discriminant is nonzero and its leading coefficient is fixed. We can therefore determine the number of zeros by examining one cubic in each root class. For three distinct real roots, choose $P=t^3-t$, giving
\begin{equation}
R=-3(t^2+1)(t^2-2t-1)(t^2+2t-1).
\end{equation}
There is one zero of $R$ in each region where $P>0$, namely $(-1,0)$ and $(1,\infty)$. For one real root, take $P=t^3+t$, yielding
\begin{equation}
R=-3(t^2-1)(t^4+6t^2+1).
\end{equation}
There is again one zero of $R$ in the region $P>0$, at $t>0$. A zero of $R$ cannot cross a simple zero $r$ of $P$, because $R(r)=3[P'(r)]^3\ne0$, and it cannot enter from infinity while the leading coefficient remains nonzero. Thus $E'''$ has at most one zero on any interval where $P>0$. Since $P$ is the squared magnitude of an exchange amplitude, it cannot have a simple root at an interior physical wave vector. Interior double roots must be considered separately.

\paragraph{Zero interlayer exchange amplitude.} If the interlayer amplitude vanishes at an interior wave vector, its squared magnitude $P$ must have a double root there, unless $P\equiv0$. For a cubic, shift that zero to $t=0$: $P(t)=Kt^2(t+a)$, with $K>0$, $a>0$ in the physical neighborhood, after reversing $t$ if needed. The quadratic perfect-square case is treated below. For the cubic,
\begin{equation}
R(t)=-3K^3t^5(t+2a).
\end{equation}
The root $t=-2a$ lies outside the physical region $t+a>0$. If the double root lies outside the physical interval, the factor $t+2a$ permits at most one zero of $E'''$ where $P>0$. For an interior double root, the exchange energy near $t=0$ contains $-\sqrt K\,|t|\sqrt{t+a}$. Its right derivative lies below its left derivative by $2\sqrt{Ka}$. The smooth exchange terms in Eq.~\eqref{eq:appE} cannot remove this downward jump. The cusp is therefore not a minimum. On either side, $E'''$ has no zero, which permits at most one isolated strict minimum by Rolle's theorem.

For a triple root, $P=Kt^3$ on its positive side and $R=-3K^3t^6$, so $E'''$ has no zero there. For quadratic $P=a+bx+cx^2$, direct differentiation gives $R=3P'(b^2-4ac)$, with at most one zero unless $P$ is a perfect square. Then $\sqrt P$ is the absolute value of a linear function and the energy is quadratic on either side of its downward cusp. Each quadratic branch has at most one isolated strict minimum. Linear or constant $P$ gives the same or a simpler restriction, and $P\equiv0$ leaves a quadratic energy with at most one isolated interior minimum.

Three isolated strict minima on one smooth interval require two intervening maxima and hence at least three zeros of $E'''$ by Rolle's theorem. The exchange energy permits at most one such zero. When $P$ vanishes within the physical interval, there can be at most one minimum on either side of the downward cusp. The optimized homogeneous spiral energy therefore has at most two distinct noncollinear minima for arbitrary real exchanges. Their stability against small fluctuations can be tested with the harmonic matrices in Appendix~\ref{app:mat}.

\section*{Data Availability Statement}
The exchange parameters and equations needed to reproduce the reported numerical results and figures are given in the article.

\begin{acknowledgments}
Work at Ames National Laboratory was supported by the U.S. Department of Energy (DOE), Office of Science, Basic Energy Sciences, Materials Sciences and Engineering Division. Ames National Laboratory is operated for the U.S. DOE by Iowa State University under Contract No. DE-AC02-07CH11358.
\end{acknowledgments}

\end{document}